\documentclass[prl,reprint,twocolumn,superscriptaddress,noshowpacs,notitlepage,longbibliography]{revtex4-1}%
\usepackage{graphicx,bm,times}
\usepackage{amsmath}
\usepackage{amsfonts}
\usepackage{amssymb}
\usepackage{color}
\usepackage{hyperref}
\hypersetup{
	colorlinks = true,
	allcolors = {blue}
}

\begin{document}
	
	\title{Orbital- and k${_\mathbf{z}}$-selective hybridisation of Se 4p and Ti 3d states\\ in the charge density wave phase of TiSe$\mathbf{_2}$}
	
	\author{Matthew D. Watson}
	\email{mdw5@st-andrews.ac.uk}
	\affiliation {SUPA, School of Physics and Astronomy, University of St. Andrews, St. Andrews KY16 9SS, United Kingdom}
	\affiliation{Diamond Light Source, Harwell Campus, Didcot, OX11 0DE, United Kingdom}
	
	\author{Oliver J. Clark}
	\author{Federico Mazzola}
\affiliation {SUPA, School of Physics and Astronomy, University of St. Andrews, St. Andrews KY16 9SS, United Kingdom}

	\author{Igor~Markovi{\'c}}
	\author{Veronika~Sunko}
\affiliation {SUPA, School of Physics and Astronomy, University of St. Andrews, St. Andrews KY16 9SS, United Kingdom}
\affiliation {Max Planck Institute for Chemical Physics of Solids, N{\"o}thnitzer Stra{\ss}e 40, 01187 Dresden, Germany}
			
	\author{Timur K. Kim}
	\affiliation{Diamond Light Source, Harwell Campus, Didcot, OX11 0DE, United Kingdom}
	
	\author{Kai Rossnagel}
	\affiliation{Institut f{\"u}r Experimentelle und Angewandte Physik, Christian-Albrechts-Universit{\"a}t zu Kiel, 24098 Kiel, Germany}
	\affiliation{Ruprecht-Haensel-Labor, Christian-Albrechts-Universität zu Kiel und Deutsches Elektronen-Synchrotron DESY, 22607 Hamburg, Germany}	
	\affiliation{Deutsches Elektronen-Synchrotron DESY, 22607 Hamburg, Germany}
	
	\author{Philip D. C. King}
	\email{philip.king@st-andrews.ac.uk}
\affiliation {SUPA, School of Physics and Astronomy, University of St. Andrews, St. Andrews KY16 9SS, United Kingdom}	
	
	\begin{abstract}
	
\noindent We revisit the enduring problem of the $2\times{}2\times{}2$ charge density wave (CDW) order in TiSe$_2$, utilising photon energy-dependent angle-resolved photoemission spectroscopy to probe the full three-dimensional high- and low-temperature electronic structure. Our measurements demonstrate how a mismatch of dimensionality between the 3D conduction bands and the quasi-2D valence bands in this system leads to a hybridisation that is strongly $k_z$-dependent. While such a momentum-selective coupling can provide the energy gain required to form the CDW, we show how additional ``passenger'' states remain, which couple only weakly to the CDW and thus dominate the low-energy physics in the ordered phase of TiSe$_2$.
		
	\end{abstract}
	\maketitle
	

Charge-density waves (CDWs), typically found in low-dimensional metallic systems, are a striking example of how coupling between the electrons and the crystal lattice can lead to novel ground states. In quasi-1D systems, such as ZrTe$_3$ and organic salts, the CDW phenomenology often closely resembles the famous Peierls instability \cite{Hoesch2009PRL,Pouget2016}. Transition metal dichalcogenides (TMDC) such as NbSe$_2$ have become prototypical examples for how CDWs can also occur in quasi-2D metals \cite{Johannes2008,	Borisenko2009,Rossnagel2011}. The starting point for understanding the CDW in the well-known TMDC TiSe$_2$ is, however, an indirect narrow-gap semiconductor. Moreover, the CDW ordering pattern and the normal state electronic structure are three-dimensional \cite{DiSalvo1976,Bianco2015}, placing the CDW in TiSe$_2$ far from the conventional picture. Some reports assign TiSe$_2$ as a realisation of the long-predicted `exciton insulator' phase, driven purely by electronic interactions \cite{Cercellier2007,Cazzaniga2012,Monney2010PRB,Monney2011PRL,Hellmann2012,Monney2012NJP,Kogar2017}. The presence of substantial lattice distortions in the ordered phase, however, also points to an important role of electron-phonon coupling,  \cite{Yoshida1980,Suzuki1985,Rossnagel2002,Rossnagel2011,Weber2011}, which may act cooperatively with electron-hole correlations to stabilise the CDW \cite{Kaneko2018,Porer2014,vanWezel2010,Monney2016tr}. Despite decades of study, the origin and nature of the CDW-like instability in TiSe$_2$ thus remains highly controversial. 

\begin{figure*}
	\centering
	\includegraphics[width=\linewidth]{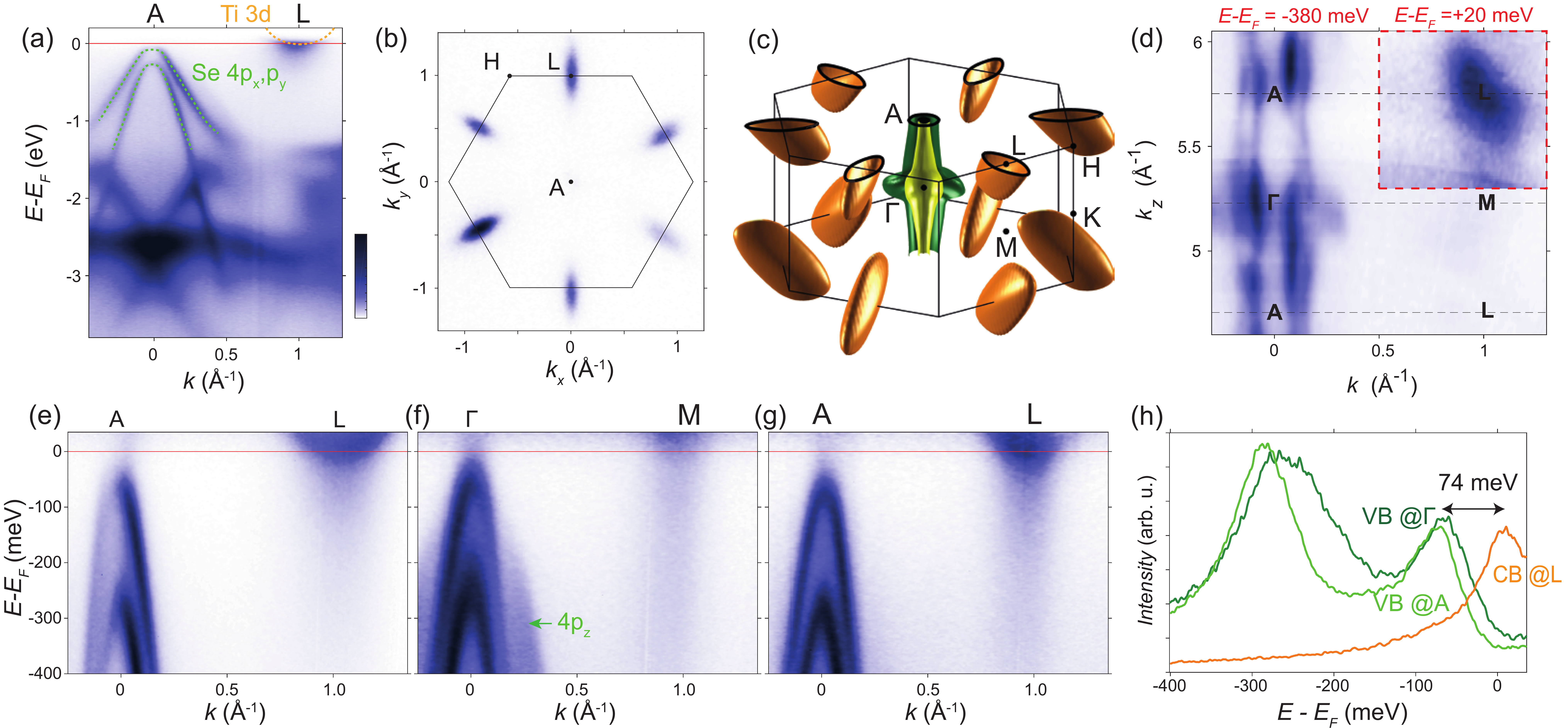}
	\caption{(a) Valence band dispersion along A-L at T = 300~K, measured at a photon energy $h\nu$ = 121 eV, in $p$-polarisation ($p$- and $s$-polarisations correspond to Linear Horizontal and Linear Vertical polarisations, respectively, and the analyser slit is aligned vertically at the I05 beamline). (b) Constant energy map at $E_F$ (Fermi surface) at T = 250 K ($h\nu$ = 121 eV, $p$-polarisation). (c) Fermi surface of TiSe$_2$ as calculated with a standard DFT code \cite{wien2k}. (d) $k_z$ map, processed from photon-energy dependent data at $E\!-\!E_F$~= -380 meV. The inset shows the conduction band at $E\!-\!E_F$~=~+20 meV from the same data set. (e-g) Dispersions at T = 300~K acquired in $s$-polarisation with $h\nu$~=~44, 95, and 119~eV, approximately corresponding to L-A-L, M-$\Gamma$-M, and L-A-L paths respectively. (h) Comparison of EDCs at high-symmetry points, indicating an indirect band gap of 74$\pm$15 meV. The data in  panels (e-h) are divided by the Fermi function to reveal the spectral weight above $E_F$.}
	\label{fig1}
\end{figure*}

In this Letter, we revisit the evolution of the electronic structure of TiSe$_2$ through the CDW transition, paying particular attention to $k_z$-dependent variations by employing detailed photon-energy dependent ARPES. This allows us to comprehensively determine the positions of the band extrema in this system, identifying a low-temperature band gap that is, surprisingly, smaller than that at high temperatures. We show that this results from a strongly orbital- and $k_z$-selective hybridisation of states involved in the $2\times{}2\times{}2$ CDW, whereby additional ``passenger" states which are not strongly hybridised dominate the low energy electronic structure in the ordered phase. 

Single crystals were grown by the iodine vapour transport method, and cleaved \textit{in-situ} (see Supplemental Material, SM \cite{SM}). ARPES measurements were performed at the I05 beamline at Diamond Light Source \cite{I05beamlinepaper}. We label high-symmetry points of the Brillouin zone according to the notation of the high temperature phase, using the starred $\Gamma^*$ notation when referring to the 2$\times{}$2$\times{}$2 low-temperature unit cell.  
 

The understanding of the CDW instability in TiSe$_2$ rests upon the details of the normal state electronic dispersions, and in particular the dimensionality of the relevant bands. Fig.~\ref{fig1}(a) shows an overview of the occupied states, with a pair of Se $4p_{x,y}$ states forming the low-energy valence bands at the Brillouin zone center, and a Ti $3d$-derived conduction band centered at each L point \cite{Chen2016,Rossnagel2002}. The conduction bands have an elliptical cross section in the $k_x-k_y$ plane, evident in a constant energy map at $E_F$ measured using a photon energy chosen to probe approximately the A-L-H plane ($k_z\approx\pi/c$, Fig.~\ref{fig1}(b)). They are, however, strongly three-dimensional. In Fig.~\ref{fig1}(e,g), the ARPES measurements at L points of the Brillouin zone show a clear electron-like conduction band, but the band disperses significantly along $k_z$ such that it is well above $E_F$ at M, in Fig.~\ref{fig1}(f). Consistent with the Fermi surface calculated from density functional theory (DFT, using the generalised gradient approximation \cite{SM}), shown in Fig.~\ref{fig1}(c), our photon energy-dependent measurements (Fig.~\ref{fig1}(d)~\cite{kz}) also indicate that this conduction band pocket substantially tilts away from the $k_z$ axis. In contrast, the uppermost valence bands, of Se $4p_{x,y}$ character, are found to be almost 2D; measurements with photon energies chosen to probe the high-symmetry $\Gamma$ and A points (Fig.~\ref{fig1}(e-g)) indicate a $k_z$ dispersion of the upper valence band of $<$10~meV (Fig.~\ref{fig1}(h)). A third valence band, of Se $4p_{z}$ character, appears only around the $\Gamma$ point and disperses strongly along $k_z$ (Fig.~\ref{fig1}(d)). This band does not play an important role in the CDW ordering and so we do not consider it in detail below.

The normal state band gap of TiSe$_2$ is a critical parameter, particularly in the context of possible excitonic effects, but previous experimental estimates have varied from a gap of 150 meV to a band overlap of 70 meV \cite{Rasch2008,Zhao2007,Kidd2002,Chen2016,Cercellier2007}. To settle this, in Fig.~1(h) we compare energy distribution curves (EDCs) of the valence bands measured at $\Gamma$ and A with the conduction band minimum at L. From this, we can confidently estimate the gap between the $\Gamma$ and L points to be 74$\pm$15~meV. Thus in stark contrast to standard DFT calculations which predict a semimetallic state with a large band overlap (SM \cite{SM}), the normal state of TiSe$_2$ is a semiconductor with a narrow indirect band gap of 74$\pm$15~meV. Crucially, however, this small magnitude of the band gap allows the valence band states to participate in the CDW-like physics by hybridising with the conduction states.

\begin{figure}[t]
	\centering
	\includegraphics[width=\linewidth]{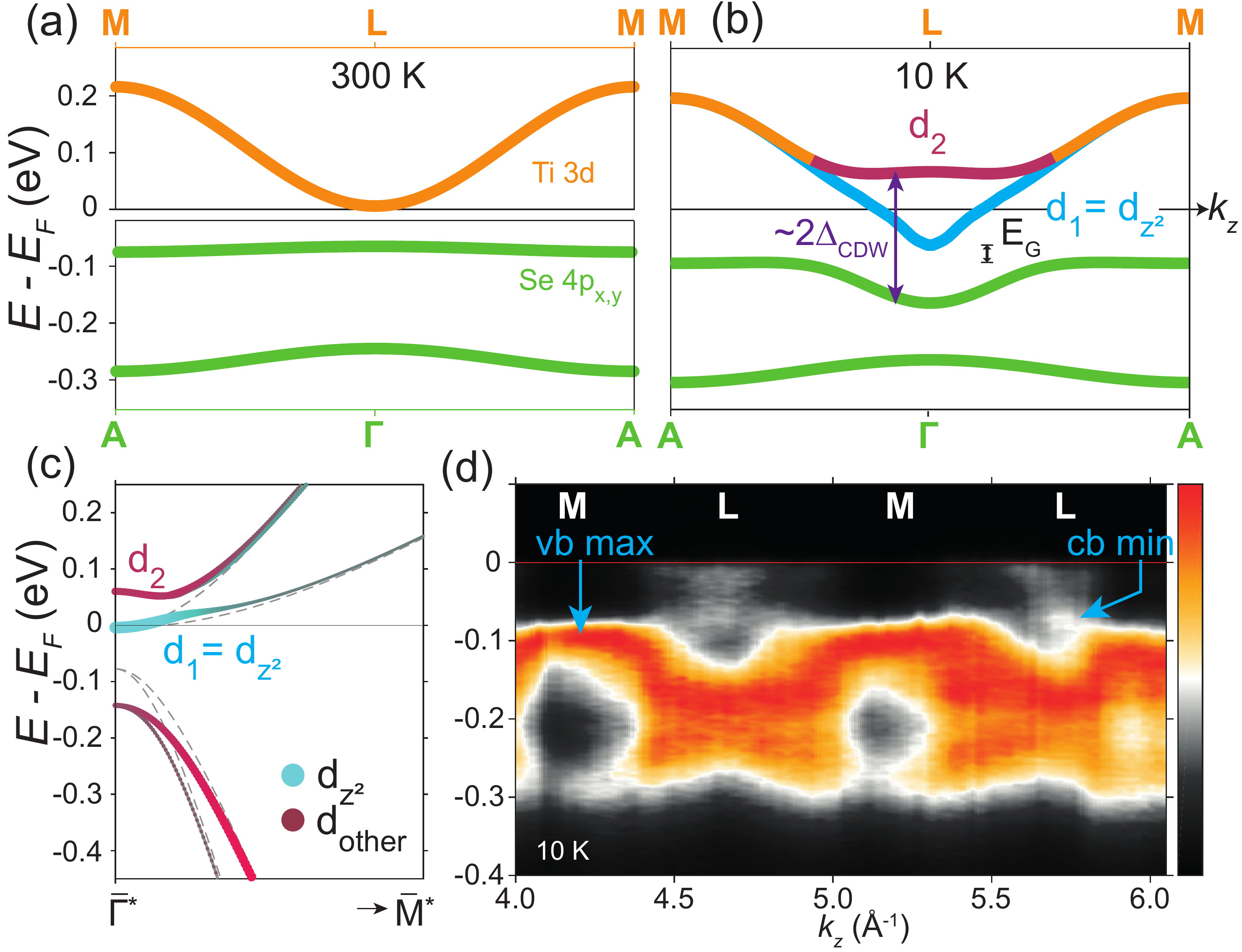}
	\caption{a) Schematic $k_z$-dispersions in the normal state. The conduction bands are plotted with a displacement of $q_{CDW}$. The $4p_z$ valence band is not included. b) Schematic $k_z$-dispersions in the 3Q CDW phase, showing $k_z$-selective band hybridisation. We do not include the backfolded bands. c) Tight-binding calculations of monolayer TiSe$_2$ showing orbital-selective band hybridisation in the CDW phase. Dashed gray lines represent band dispersions at infinitesimal distortion. Color and weight of lines project the difference between the Ti $3d_{z^2}$ character and other $3d$ components; see SM for further plots and details. d) Experimental $k_z$-dispersion along M-L-M in the CDW phase. The intensity is dominated by bright backfolded valence bands, which display characteristic dips around L points. Conduction pockets are also observed, centered at L points.} 
	\label{fig2}
\end{figure}

\begin{figure*}
	\centering
	\includegraphics[width=0.96\linewidth]{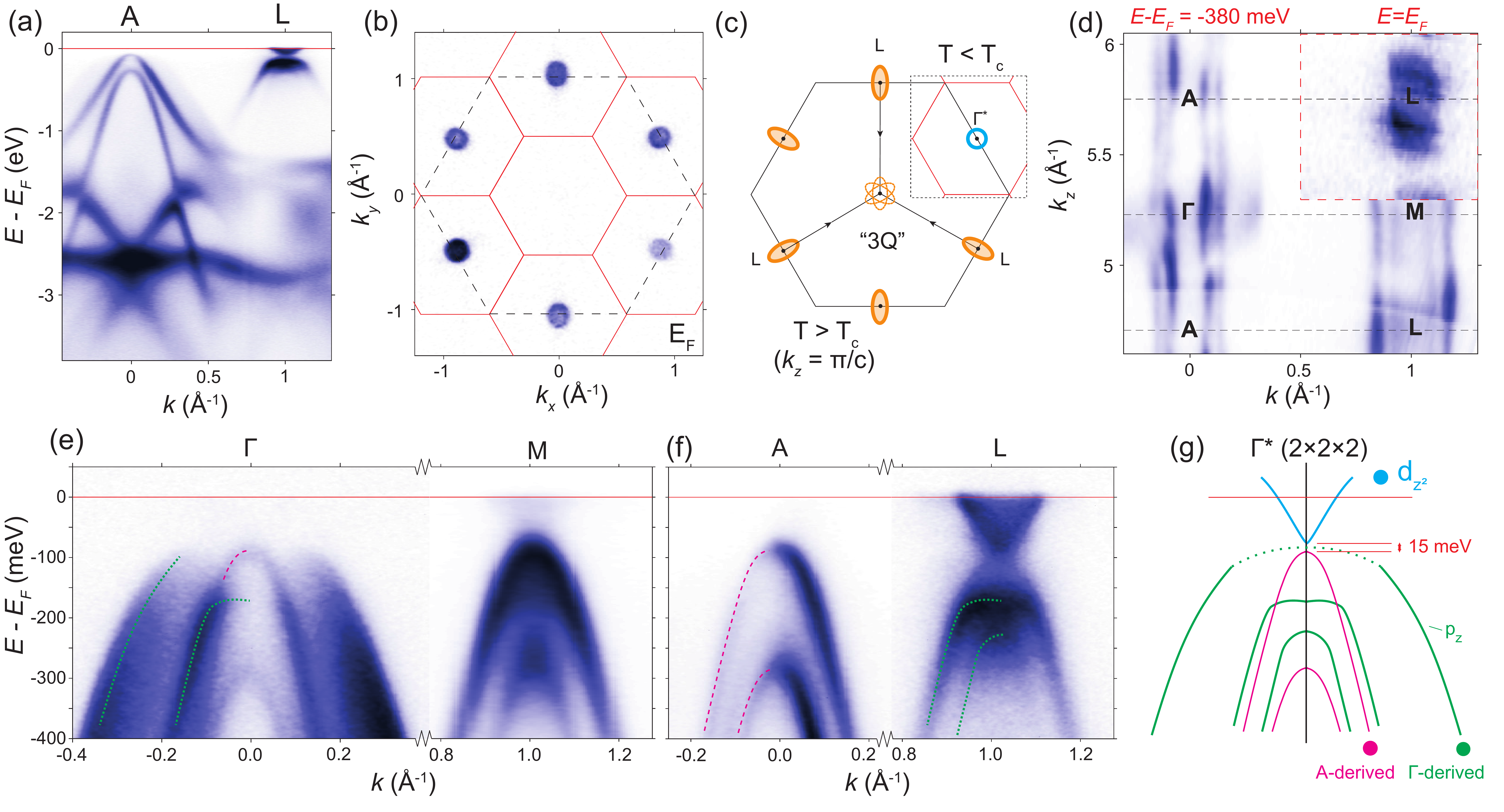}
	\caption{a) Valence band dispersion along A-L at 10~K ($h\nu$~=~121 eV, $p$-polarisation). b) Fermi surface map, showing near-circular pockets at the L points. c) In-plane construction of the 3Q ordered phase; backfolded bands are dashed, the reconstructed low-temperature Fermi surface is shown inset. d) Low-temperature $k_z$ map, equivalent to Fig.~\ref{fig1}(d), now showing backfolded valence bands, and the Fermi surface in the inset (top-right). e,f) High symmetry dispersions along $\Gamma$-M (31 eV, 100 eV) and A-L (44 eV, 121 eV, all $s$-polarisation) respectively. g) Schematic band structure at $\Gamma^*$ of the $2\times{}2\times{}2$ Brillouin zone, with an estimated band gap of $\sim$15 meV.}
	\label{fig3}
\end{figure*}

The mismatched dimensionality of the conduction and valence bands, however, has a major influence on shaping the electronic structure in the ordered state of TiSe$_2$. In Fig.~\ref{fig2}(a) we show the normal state conduction and valence band dispersions along $k_z$. These are based on the data in Fig.~\ref{fig1}, except for the $k_z$ dispersion of the conduction band, for which we use a bandwidth of $\sim$200~meV from DFT calculations. In the CDW phase, the valence and conduction bands hybridise according to an interaction $\Delta_{CDW}(\mathbf{k})$. While this depends on the microscopics of the interaction terms, a reasonable assumption for $\Delta_{CDW}(\mathbf{k})$ is a broadly-peaked function centered around $\mathbf{q}_{CDW}=(\pi,0,\pi)=\mathrm{L}$, where phonon modes are known to soften at the CDW transition of TiSe$_2$~\cite{Weber2011,Holt2001}.

Anticipating the hybridisations allowed by this interaction term, we plot the bands in Fig.~\ref{fig2}(a) offset by $\mathbf{q}_{CDW}$. In this way, the valence band maximum at $\Gamma$ and the conduction band minimum at L coincide. They can therefore be expected to be strongly hybridised at the CDW transition: as illustrated in Fig.~\ref{fig2}(b), the valence bands from $\Gamma$ and the branch of the conduction band labelled $d_{2}$ are significantly repelled from the Fermi level, opening up an energy gap on a scale of $\sim2\Delta_{CDW}$. The unoccupied $d_2$ states are not detectable by ARPES, but are consistent with inter-band transitions observed by resonant inelastic x-ray scattering \cite{Monney2012PRL}. If these were the only states in the system, this energy scale would also correspond to the low temperature band gap. There are, however, other states to consider. 

First, the upper valence band at the A point cannot hybridise significantly with the conduction band states, since the pronounced $k_z$ dispersion of the latter renders it at inaccessibly high energies at M (Fig.~\ref{fig2}(a,b)). Therefore the dispersion of the valence bands derived from the A point of the high temperature phase will remain largely unchanged through the CDW transition, except for an overall shift of the chemical potential \cite{SM}. The band hybridisation is thus strongly $k_z$-dependent.

Moreover, the different orbital components of the conduction band couple inequivalently to the CDW order. The $2\times{}2\times{}2$ lattice distortion in TiSe$_2$ is a 3Q ordering, corresponding to a superposition of the atomic displacements associated with three softened L-point phonon modes \cite{Bianco2015}. Thus, electron pockets from three L points are backfolded to the $\Gamma^*$ point of the reconstructed Brillouin zone (Fig.~\ref{fig3}(c)), allowing hybridisation with the valence bands. To clarify the form of these hybridisations, we consider a simplified tight-binding analysis for a monolayer of TiSe$_2$ without spin-orbit coupling \cite{Kaneko2018}, explained in depth in the SM \cite{SM}. We implement a $2\times{}2$ CDW by considering only the Ti displacements, and rescaling the Ti-Se $d-p$ hoppings according to the modified bond lengths in the ordered phase. The resulting band dispersions (Fig.~\ref{fig2}(c)) show a structure similar to \textit{ab-initio} calculations \cite{Bianco2015,Hellgren2017}, where the three backfolded conduction bands split into a doublet ($d_2$) and singlet ($d_1$) branch at $\overline{\Gamma}^*$. The valence bands hybridise strongly only with the $d_2$ branch, while the $d_1$ branch remains unhybridised and is not pushed away from the Fermi level. Orbital-projections reveal that this dichotomy is due to an orbital-selectivity of the hybridisation, with the unhybridised $d_1$ branch corresponding exactly to the $d_{z^2}$ orbital projection of the conduction bands, as defined in the crystalographic reference frame. Since the atomic displacements are in-plane only~\cite{DiSalvo1976}, the extra hybridisation terms associated with the CDW distortion cancel at $\overline{\Gamma}^*$ for the d$_{z^2}$ orbital. A similar orbital selectivity should occur for bulk TiSe$_2$: as well as the $k_z$ selectivity of the $d_2$-branch hybridisation introduced above, an additional unhybridised $d_1$ branch will remain, shown schematically in Fig.~\ref{fig2}(b).

To verify the above considerations,  we search for two indicative spectroscopic signatures. The first is that the upper valence band at low temperatures should show characteristic dips along $k_z$, which can be approximately considered as the inverse profile of the interaction $\Delta_{CDW}({k_z})$. Such dips are evident in measurements of the $k_z$ dispersion of the bright backfolded valence bands along M-L-M in the low-temperature phase ($T=10$~K, Fig.~\ref{fig2}(d)), very similar to the schematic in Fig.~\ref{fig2}(b). The uppermost valence band dips by $\Delta_{CDW} \approx$ 100 meV around L points, where primarily the strongly hybridised valence bands derived from $\Gamma$ are being observed, but recovers at M points where the more weakly hybridised valence bands derived from the A point are observed, confirming $k_z$-selective hybridisation. 
       
Second, we consider the relative spectral weight of bands that are backfolded by the periodic lattice distortion. As evident in Fig.~\ref{fig3}(a), low-temperature measurements show prominent backfolding of valence bands at the L point. This is a famous feature of ARPES spectra in the CDW phase of TiSe$_2$~\cite{Pillo2000,Rossnagel2002,Kidd2002,Cercellier2007,Monney2010PRB,Chen2016,Ghafari2017,Rossnagel2011}, where the backfolded intensity is comparable with the original valence bands at the $\Gamma$ point, indicating a strong involvement of these states in the CDW. In contrast, the conduction band appears brightly at the L point, but backfolded copies of this band are not clearly apparent in the spectra measured at $\Gamma$, consistent with previous reports~\cite{Cercellier2007,Monney2009PRB}. In fact, careful measurements at the $\Gamma$ point (see Fig.~SM1 in SM) do reveal a replica of the conduction band, but with extremely weak spectral weight, showing a striking asymmetry with the remarkably bright intensity of the backfolded valence bands. This is entirely consistent with the assignment of the occupied electron band in the low-temperature phase as the branch with $d_{z^2}$ orbital character which remains unhybridised, since it hardly couples to the new periodicity and is thus only very weakly backfolded. 

The measurements and analysis above thus demonstrate that the low-energy physics in the ordered state of TiSe$_2$ is actually dominated by states which do not couple strongly to the CDW. TiSe$_2$ crystals typically exhibit a slight unintentional $n$-doping~\cite{DiSalvo1976,Rossnagel2002,Huang2017,carrierdensity}. Despite the large band shifts associated with the CDW, TiSe$_2$ therefore retains a well-defined ungapped Fermi surface at low temperature (Fig.~\ref{fig3}(b)), derived from the unhybridised $d_1$ branch. Despite the weak coupling of the states which form this Fermi surface to the periodic lattice distortion, they must still respect the symmetry of the low-temperature phase, where $\Gamma$, M, A and L points all become formally equivalent to $\Gamma^*$ of the $2\times{}2\times{}2$ Brillouin zone. The low temperature conduction band must adhere to $\bar{3}$ symmetry,  explaining their almost circular in-plane dispersion observed experimentally, in contrast to the twofold-symmetric elliptical pockets in the high-temperature phase (Fig.~\ref{fig3}(b,c)). Moreover, the slanting of the Fermi pockets away from the $k_z$ axis seen in the high-temperature phase (Fig.~\ref{fig1}(c,d)) is forbidden at low temperatures (Fig.~\ref{fig3}(d)). The conduction bands also develop a steeper dispersion at low temperatures, consistent with recent first-principles calculations~\cite{Hellgren2017}.

 There are five low-energy valence bands centered at each $\Gamma^*$ point: three derived from the high-temperature $\Gamma$ point (one of which is the $p_z$ state) and two derived from the A point \cite{Bianco2015}. These cannot all be observed simultaneously in one geometry, since the spectral weight as seen by ARPES typically follows the projection of the states onto the high-temperature zone \cite{Rossnagel2011}. However, by combining the dispersions observed at different photon energies in Fig.~\ref{fig3}(e,f), we can identify all five, represented schematically in Fig.~\ref{fig3}(g). The upper valence band derived from the $\Gamma$ point is flattened and hybridises so strongly that in fact it is more clearly observed backfolded at the L point in Fig.~\ref{fig3}(f). In contrast, due to the $k_z$-selective hybridisation, the A-point dispersion in Fig.~\ref{fig3}(f) is very similar to the equivalent measurement at high temperature (Fig.~\ref{fig1}(e)) except that the bands are sharper and there is an overall shift of the chemical potential. 
 
Our measurements in Fig.~\ref{fig3}(e-g) allow us to estimate a low-temperature band gap of only $\sim$15 meV - much smaller than the 74 meV band gap at high temperatures \cite{lowtbandgap}. This is surprising, since second-order electronic phase transitions typically involve opening up or increasing energy gaps, but the ideas of $k_z$- and orbital-selective hybridisations give insight. The strongly-hybridised states, originating from the $\Gamma$ and L points, are shifted away from $E_F$ by $\pm\Delta_{CDW}\sim$~100~meV, a large energy scale consistent with the high $T_c\approx{}$200~K. However, the band gap in the CDW phase is between two other low-energy states: the A-derived valence band and the unhybridised Ti-$d_{z^2}$ conduction band branch. The energy gap between these passenger states is decoupled from the main hybridisation that drives the CDW, and is, therefore, not required to increase in the CDW phase. The decreased gap between these states at low temperatures compared with the normal state remains a little enigmatic. We speculate that a modified out-of-plane hopping of the $d_{z^2}$ passenger-state conduction band may play a role, as well as other possible many-particle effects \cite{SM}. In any case, the key concept is that the overall band gap in the CDW phase is essentially decoupled from the strength of the ordering.

The observation of a smaller band gap in the ordered phase is not easy to reconcile with a purely excitonic mechanism for the phase transition. Moreover, we have observed an ungapped Fermi surface and a near-continuum of single particle excitations across the energy scale of $\pm{}\Delta_{CDW}$; the existence of these low-energy passenger states would presumably promote the decay of excitons. Nevertheless, the relatively low free carrier density from these passenger states may be insufficient to completely screen the Coulomb interaction between electrons and holes \cite{Porer2014}, allowing for excitonic correlations to still play an assistive role in the CDW ordering \cite{vanWezel2010,Porer2014,Watanabe2015,Kaneko2018,Monney2016tr}. Irrespective of the precise microscopic origin of the instability, its underlying driving force is an electronic energy gain from band hybridisation. Our measurements highlight how momentum and orbital selectivity can act to decouple such energy gain from the low-energy physics of the ordered state. This is of crucial importance for understanding not only the enigmatic CDW state in TiSe$_2$, but also charge ordering instabilities in multi-orbital systems in general.


\begin{acknowledgments}
	\section{acknowledgments}
	We thank Andrew Hunter, Deepnarayan Biswas, Akhil Rajan, Kaycee Underwood, Kerstin Hanff and Moritz Hoesch for technical support and useful discussions. We thank Diamond Light Source for access to Beamline I05 (Proposal Nos. SI19771-1, NT18555-1, and SI16262-1) that contributed to the results presented here. We gratefully acknowledge support from The Leverhulme Trust (Grant Nos. RL-2016-006 and PLP-2015-144) and The Royal Society. V.S. and O.J.C. acknowledge EPSRC for PhD studentship support through Grant Nos.~EP/L015110/1 and EP/K503162/1.  I.M. acknowledges PhD studentship support from the IMPRS for the Chemistry and Physics of Quantum Materials. The research data supporting this publication can be accessed at Ref.~\cite{datarepository}.
\end{acknowledgments}


\begin{thebibliography}{41}
\bibitem [{kz()}]{kz}%
	\BibitemOpen
	\href@noop {} {}\bibinfo {note} {{To convert photon energies into $k_z$, we
			employ the standard free-electron final state approximation, using an inner
			potential of 13.5~eV.}}\BibitemShut {Stop}%
\bibitem [{car()}]{carrierdensity}%
	\BibitemOpen
	\href@noop {} {}\bibinfo {note} {We estimate the carrier density here to be
		$\approx\!3\times{}10^{19}$~cm$^{-3}$.}\BibitemShut {Stop}%
\bibitem [{low()}]{lowtbandgap}%
	\BibitemOpen
	\href@noop {} {}\bibinfo {note} {{In fact 15 meV is an upper limit on the low
			temperature band gap, since we were unable to pinpoint the exact top of the
			$p_z$ band dispersion}}\BibitemShut {NoStop}%
\end{thebibliography}

%

\clearpage

\onecolumngrid

\section{Supplemental Material}

\subsection{Sample growth and characterisation}

Single crystals of 1T-TiSe$_2$ were grown by the standard iodine chemical vapor transport method. A near-stoichiometric mixture of high-purity Ti and Se with a slight Se excess was placed in a quartz ampoule together with iodine (5 mg/cm$^3$) as transport agent; the ampoule was sealed and heated in a four-zone furnace under a temperature gradient of 665-555$^\circ{}$C for 600 h. Typical crystal dimensions were about 3x3x0.15 mm$^3$. The samples display a CDW at T$_c\approx$~200~K, as confirmed by resistivity and temperature-dependent ARPES. 

In Fig. 3 of the main text, it is clear that at low temperature there is a partially occupied electron band, indicating a finite n-type carrier density (i.e. a departure from exact sample stoichiometry). The exact stoichiometry was found to vary a little between samples, with some samples exhibiting slightly larger electron pockets at low temperature, and correspondingly showing rigid band shifts of up to $\sim$20~meV compared with the samples used for the data in the main text. However the CDW physics seems to be robust to variations in stoichiometry on this scale. At high temperature (Fig. 1(e,g)), the conduction band minimum is observed to lie slightly above the chemical potential. This is due to the influence of the elevated temperature: at 300~K, the tail of the Fermi-Dirac state occupation function extends well into the conduction band, and so there is still a significant population of electrons at high temperature. Thus our data are consistent with a slight $n$-doping at both high and low temperature.  

\subsection{Details of DFT calculation}

The Fermi surface in Fig.~1(c) of the main text was calculated using the Wien2k code. The Generalised Gradient Approximation was used (PBE-GGA), and spin-orbit coupling was included. The calculation was based on the experimental lattice positions in the undistorted phase, taken from submission number 108739 of the Inorganic Crystal Structure Database. A large $k$-grid (51$\times{}$51$\times{}$26) was used to allow for accurate 3D plotting of the Fermi surface.

This ``standard" DFT calculation yields a semimetallic solution, as indicated by the presence of a compensated Fermi surface, with both hole and electron-like bands. It is not possible to be precise on the calculated band overlap (negative band gap) since the calculation introduces unphysical band hybridisations above $E_F$ at the $\Gamma$ point, but it is on the order of -600 meV. While we find the calculation to be a useful indicator of the dimensionality of the bands, it seriously overestimates the band overlap, since the experiments show a narrow band gap of +74 meV. The recent study of Hellgren \textit{et al.} \cite{Hellgren2017} demonstrated that hybrid functionals can be used to substantially reduce this overestimated band overlap, following earlier work by Bianco \textit{et al} \cite{Bianco2015} who added a repulsive term on the more localised Ti $d$ sites by the LDA+U method to achieve a similar outcome. We refer the reader to these papers (and references therein) for further discussion of the \textit{ab-initio} perspective on TiSe$_2$.

\clearpage

\subsection{Backfolding of the conduction band}
\begin{figure}[h]
	\centering
	\includegraphics[width=0.6\linewidth]{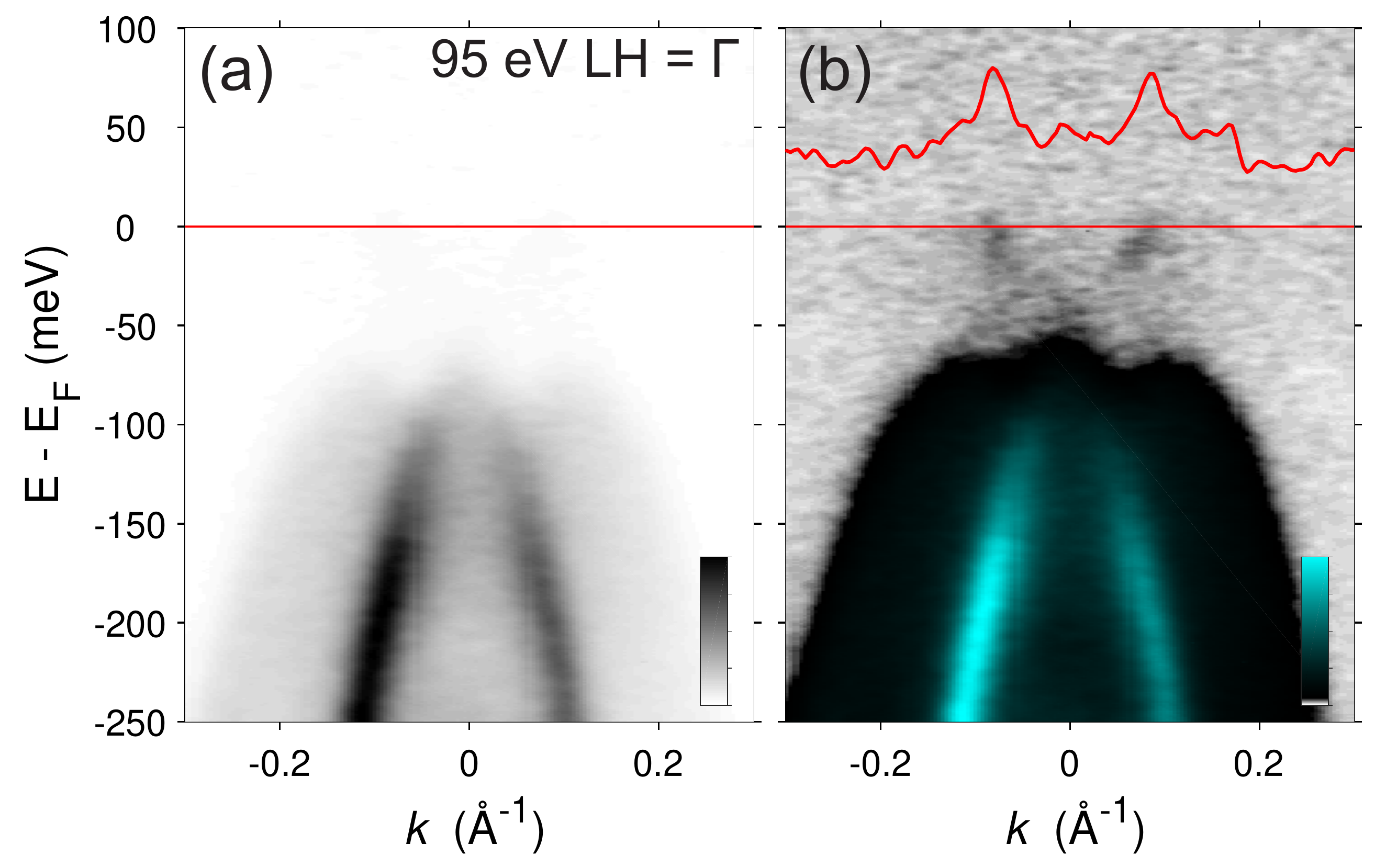}
	\caption{(a) Valence band dispersion, measured using $p$-polarisation at 95 eV, corresponding to a $\Gamma$ point. Note that this is the opposite polarisation to most of the valence band data shown in the main paper, which were mostly measured in $s$-polarisation. (b) The same data set, but with a significant overexposure of the colormap, which allows the visual observation of the very weak backfolded conduction band. The MDC at the Fermi level is also plotted, showing weak peaks at $\pm{}k_F$ of the electron band.}
	\label{fig:smfig1}
\end{figure}

In the main text, we noted that the conduction band can be observed backfolded, but with very weak intensity in ARPES. To understand this effect, we should first remark that although the bandstructure, i.e. the dispersion of the poles in the Green's function, must obey the symmetry and periodicity of the 2$\times{}$2$\times{}$2 unit cell, the \textit{spectral weight} as probed by ARPES tends to mainly follow the periodicity of the original cell. The spectral weight of ``backfolded" features is then typically proportional to the magnitude of their coupling to the new periodicity, though modified matrix elements may also affect the measurements. Theoretically, this problem is treated by projecting (or ``unfolding") the supercell wavefunctions back onto the basis of the original cell \cite{Bianco2015}. The experimental consequence is that one cannot simply expect to measure all the folded dispersions corresponding to the $\Gamma^*$ point in a single geometry; rather one needs to build a picture by measuring at each of the $\Gamma$,A,M, and L points of the high-temperature cell, and identify the unique dispersions. The low temperature conduction band is not a strongly hybridised state, and derives from the $d_{z^2}$ orbital without significant hybridisation in the CDW phase. Since this conduction band is therefore only weakly coupled to the CDW periodicity, it is observed strongly at the L point (where the conduction band states exist in the undistorted phase) but with only very weak intensity at the $\Gamma$ point, even though both are formally $\Gamma^*$ points of the ordered phase. This asymmetry of spectral weight has been also found in \textit{ab-initio} \cite{Bianco2015} and tight-binding modelling \cite{Kaneko2018}. 

\
\clearpage

\subsection{Orbital character of conduction bands: tight-binding model calculations}

\begin{figure}[h]
	\centering
	\includegraphics[width=\linewidth]{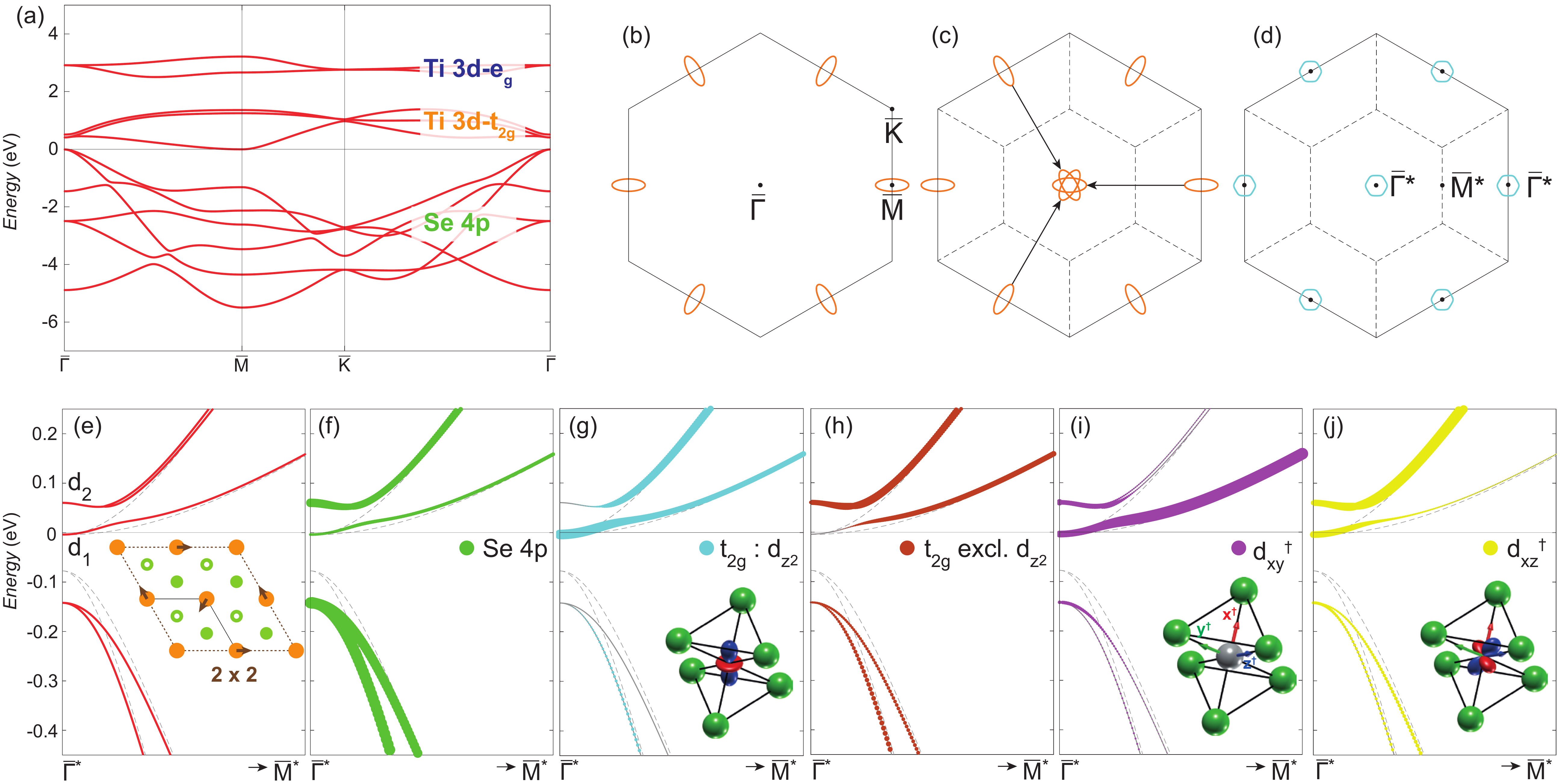}
	\caption{Monolayer TiSe$_2$ tight-binding simulations. (a) Bandstructure in the undistorted phase. Distinct bandwidths arising from primarily the Se $4p$, Ti $3d:t_{2g}$ and $3d:e_g$ orbitals are identified. (b) Fermi surface in the undistorted phase. (c) Backfolding of the Fermi surface in the CDW phase at infinitesimal distortion. (d) Calculated low-temperature Fermi surface for a realistic value of the CDW distortion/interaction. (e) Detailed band dispersions at the $\bar{\Gamma}^*$ point of the reconstructed phase, identifying the strongly hybridised doubly-degenerate $d_2$ branch and the singly-degenerate $d_1$ unhybridised branch. Dashed lines correspond to infinitesimal distortion (i.e. dispersions are shown backfolded, but not hybridised). Inset shows the 2$\times{}$2 structure of the distorted phase. Ti and Se atoms are indicated as orange and green icons, respectively. Arrows indicate the Ti displacements considered. In our tight-binding analysis, we neglect the smaller displacements of the heavier Se atoms. (f-j) Orbital character projections. In (f), one can see the extra Se weight appearing on the upper, hybridised branch of the conduction band, while the lower, unhybridised branch, does not gain Se weight. (g) The $d_1$ conduction band is clearly identified with the $d_{z^2}$ orbital, whereas in (h) the doubly-degenerate $d_2$ branch corresponds to the projections of $t_{2g}$ orthogonal to $d_{z^2}$. (i,j) Moving away from the high-symmetry point and the hybridised region, the natural orbital basis are the rotated basis functions where $x^\dagger{}$,$y^\dagger{}$,$z^\dagger{}$ correspond to the idealised octahedral environment of Ti (colored arrows in inset). }
	\label{fig:smtb}
\end{figure}

\subsubsection{Monolayer tight binding model: the normal phase}  

We use the model recently developed by Kaneko \textit{et al.} \cite{Kaneko2018}. This model was developed for monolayer TiSe$_2$, and therefore does not capture the $k_z$-dependent physics we report in the main text; however it provides a useful simplification to allow discussion of structure and orbital content of the conduction bands. Kaneko \textit{et al.} \cite{Kaneko2018} fitted the band dispersions obtained from DFT calculations of the monolayer to obtain their Slater-Koster parameters. However, since their parameters give a small band overlap (negative band gap), we adjust the $p$-orbital energy down by 100 meV in order to yield a band gap comparable with the bulk value, keeping other parameters the same as in Ref.~\cite{Kaneko2018}. We do not claim this parameter set is optimized compared with the experimental dispersions for either bulk or monolayer TiSe$_2$, only that the model is sufficiently realistic to make statements regarding the orbital character of the relevant low-energy states of TiSe$_2$. We use a barred notation for high-symmetry points ($\bar{\Gamma}$ etc) to emphasize that this is a monolayer calculation, distinct from the 3D bulk data. The starred $\bar{\Gamma}^*$ notation refers to the unit cell of the reconstructed $2\times{}2$ Brillouin zone, in keeping with the notation of the main text. 

Considering first the undistorted phase in Fig.~\ref{fig:smtb}(a), we find a clear hierarchy of bandwidths. The Se $4p$ states form strongly-dispersive valence bands. Notably, the $p_z$ orbital which disperses strongly along $k_z$ in the bulk is confined in the monolayer case, thus only the $p_x,p_y$ valence bands appear at low energies. Moreover our tight-binding implementation does not include spin-orbit coupling, meaning that these states form a doublet at the $\bar{\Gamma}$ point, instead of hosting a splitting of $>200$~meV as observed experimentally (see e.g. Fig.~1(f) of the main text). The Ti $3d$ orbitals form relatively weakly dispersing distinct $t_{2g}$ and $e_{g}$ manifolds, split by the crystal field environment of the octahedral coordination of Ti. The $t_{2g}$ bands are further split at the $\bar{\Gamma}$ point due to trigonal distortions of the octahedral environment in the 1T phase, but this is only a small effect. The term $t_{2g}$ in this context corresponds to the $d^\dagger{}_{xz},d^\dagger{}_{yz},d^\dagger{}_{xy}$ orbitals, defined in a rotated reference frame corresponding to the idealised octahedral axes, i.e. the $x^\dagger{}$,~$y^\dagger{}$,~$z^\dagger{}$ axes point approximately along each of the Ti-Se bond directions (insets of Fig.~\ref{fig:smtb}(i,j)). Thus the $t_{2g}$ orbitals determine the Fermi surface shown in Fig.~\ref{fig:smtb}(b), with the chemical potential set to slight electron-doping as is experimentally the case. These Fermi surfaces form ellipses centered at the $\mathrm{\bar{M}}$ points.

\subsubsection{Implementation of the $2\times{}2$ CDW phase} We implement the CDW distortions in a simplified manner, where only the Ti displacements are considered, and the nearest-neighbour $p-d$ hopping terms are simply rescaled by the ratio of the new bond length in the $2\times{}2$ phase to that in the undistorted phase. The actual atomic positions are not displaced for the calculation, and the non-radial components of the displacements are ignored. Although this is therefore only a first approximation, it can be justified by the fact that we obtain similar results to the complete but technically complex implementation performed by Kaneko \textit{et al.} \cite{Kaneko2018}. 

In the distorted phase, three copies of the elliptical Fermi surface are backfolded to the $\bar{\Gamma}$ point, shown in Fig.~\ref{fig:smtb}(c). Thus these bands would be triply-degenerate at the $\bar{\Gamma}^*$ point of the distorted phase at infinitesimal displacement. However, when switching on the hybridisation with the valence bands due to the interaction term introduced by the symmetry-breaking atomic displacements, this triplet splits into a doublet which strongly hybridises with the valence bands ($d_2$), and a singlet representation which is unaffected by the CDW distortion ($d_1$), as seen in in Fig.~\ref{fig:smtb}(e). Thus the Fermi surface left in the CDW phase, shown in Fig.~\ref{fig:smtb}(d) derives from the non-hybridised $d_1$ state.

\subsubsection{Orbital character of the hybridised and non-hybridised electron bands} 
It is useful to break down the orbital content of these separate dispersions, and in Fig.~\ref{fig:smtb}(g) we show that the singlet unhybridised state derives from the $d_{z^2}$ orbital. Care must be taken on the nomenclature; here the $d_{z^2}$ orbital corresponds to the crystallographic reference frame where the $d_{z^2}$ orbital points out of the plane, i.e. not the rotated reference frame. Formally it can be shown that $d_{z^2}$ is equivalent to the symmetric linear combination of $d^\dagger{}_{xz}$,$d^\dagger{}_{yz}$, and $d^\dagger{}_{xy}$. These results therefore demonstrate that the component of the $t_{2g}$ manifold which projects onto the $d_{z^2}$ orbital does not hybridise with the Se $p$-states at $\bar{\Gamma}$ (Fig.~\ref{fig:smtb}(f,g)). This behavior is related to the high-symmetry of the $d_{z^2}$ orbital in the CDW phase; since the atomic displacements are in-plane only \cite{DiSalvo1976}, the extra hybridisation terms associated with the CDW distortion cancel as a first approximation for the $d_{z^2}$ orbital, at least exactly at $\bar{\Gamma}^*$. Thus as the hybridisation is switched on, this singlet branch remains, while the doublet conduction band, corresponding to the non-$d_{z^2}$ projections of the $t_{2g}$ orbitals (Fig.~\ref{fig:smtb}(h)), hybridises with the valence bands, which are mutually repelled from the Fermi level. Away from the $\bar{\Gamma}^*$ point and the region of hybridisation, the bands revert to their dispersions in the undistorted phase, where again the natural orbital basis is that of the rotated reference frame (Fig.~\ref{fig:smtb}(i,j)). 

A singlet representation that remains essentially unhybridised also appears in DFT calculations in the CDW phase \cite{Bianco2015,Hellgren2017}. This band was particularly noted by Hellgren \textit{et al.} \cite{Hellgren2017}, who assigned it as having ``dominant $d_{z^2}$ character derived from the Ti atom in the supercell that does not move with the distortion". However in our analysis, the essential ingredient is the $d_{z^2}$ orbital character of this band, as it's character has contributions from all the Ti atoms, not just the Ti site which does not move. Thus, we find an orbital-selective hybridisation, rather than any distinction based on the atomic site.

\subsubsection{Comparison to bulk data in the main text}

Although this tight-binding analysis gives us confidence on the identification of the conduction band, experimentally we find that in bulk TiSe$_2$ this band appears to sink to higher binding energies at low temperatures, causing the total band gap of the system to reduce. In our model this feature is not reproduced; the energy of the singlet $d_{z^2}$ band at $\bar{\Gamma}^*$ simply remains approximately constant as a function of increasing hybridisation. A relevant note for understanding the experimental case is that band gaps in semiconductors are generally temperature-dependent, and these variations would be relatively pronounced in a narrow band gap system. A more specific explanation could be that since the low temperature conduction band is constituted from the $d_{z^2}$ orbital only, with lobes of the wavefunction pointing out of the plane, there may be an associated change in the inter-layer hopping for this band in the bulk case, not captured in our monolayer tight-binding model. This is, however, a subtle feature, and it remains to be seen if any \textit{ab-initio} or 3D tight-binding analysis can account for it. %

\end{document}